\documentclass[onecolumn,preprint,numbers,amsmath,amssymb]{revtex4}
\usepackage{graphicx}
\usepackage{dcolumn}
\usepackage{bm}
\begin{document}

\title{Anomalous Magnetic Susceptibility and Hall Effect from Valley Degrees of Freedom}
\author{Tianyi Cai$^{1,2}$, Wang Yao$^{1,3}$,
Shengyuan A. Yang$^1$, Junren Shi$^2$ and Qian Niu$^{1} $}%
\affiliation{ $^1$Department of Physics, the University of Texas at Austin, Austin, Texas 78712, USA \\
$^2$Institute of Physics, Chinese Academy of Sciences and ICQS,
Beijing 100080, China \\
$^3$Department of Physics, and Center for theoretical and
computational Physics, the University of Hongkong, Hongkong }
% Force line breaks with \\
\par

\begin{abstract}
We study the magnetic and transport properties of epitaxial
graphene films in this letter. We predict enhanced signal of
magnetic susceptibility and relate it to the intrinsic valley
magnetic moments. There is also an anomalous contribution to the
ordinary Hall effect, which is due to the valley dependent Berry
phase or valley-orbit coupling.
\end{abstract}

 \maketitle

Graphene, a monolayer carbon honeycomb lattice, has extraordinary
electronic properties[1,2]. Its conduction band and valence band
form conically shaped valleys, touching at two nonequivalent
corners of the hexagonal Brillouin zone called Dirac points. For
optical and electronic application, it is desirable to make
graphene semiconducting by opening a gap. A gap can be produced
either by size quantization effects in graphene nano-ribbons[3],
or by inversion symmetry breaking for epitaxial graphene grown on
the top of crystals with matching lattices such as BN or SiC [4].
For the latter case, a significant energy gap has been predicted
from ab initio calculation[5], although the experimental evidence
for the gap is still under debate[6]. Experimentally, epitaxial
graphene is relatively disordered and heavily doped because of
close contact with the substrate. However, interesting physics can
still manifest due to robust topological effects associated with
the valley degree of freedom.

In this Letter, we present our studies of magnetic and transport
properties of epitaxial graphene systems. It is assumed that the
interaction between the graphene layer and the substrate breaks
the sublattice symmetry, which opens up a band gap at Dirac
points[6]. We predict enhanced signals of magnetic susceptibility
and anomalous contribution to the ordinary Hall effect. It is
demonstrated that these extroadinary effects are due to the
special valley degree of freedom in the system. Furthermore, we
argue that both effects are robust against disorder, hence should
be readily detected in experiment.

\textit{Magnetic susceptibility.} Recent studies on the orbital
magnetism of 2D massless Dirac fermions have shown that the unique
Landau-level structure gives rise to a singular diamagnetism[7].
The susceptibility becomes highly diamagnetic when the chemical
potential is at band-touching point ($\mu$=0 eV) where the density
of states vanishes,
\begin{equation}
\chi _0 \left( \mu  \right) =  - \frac{{e^2 a^2 t^2 }}{{4\pi \hbar
^2 }}\frac{\beta}{(e^{\beta\mu/2}+e^{-\beta\mu/2})^2}.
\end{equation}
Here, \textit{t} is the nearest neighbor hopping energy,
\textit{a} is the lattice constant and $\beta=1/k_BT$. $T$ is the
temperature and $k_B$ is the Boltzman constant. It is still
mysterious about the mechanism of such a singular behavior. A
natural question is what happens to the susceptibility when a
finite gap opens at the Dirac points.

With a staggered sublattice potential, the low-energy effective
Hamiltonian near the Dirac points is given by[8,9]
\begin{eqnarray*}
H &=&\left[
\begin{array}{cc}
\frac{\Delta }{2} & v_{0}\left( \Pi_{x}\tau _{z}-i\Pi_{y}\right) \\
v_{0}\left( \Pi_{x}\tau _{z}+i\Pi_{y}\right) & -\frac{\Delta }{2}%
\end{array}%
\right],
\end{eqnarray*}
where  $v_0=\sqrt{3} at/2$ and $\Pi\equiv \textbf{k}+e\textbf{A}$
is the reduced momentum operator in a magnetic field. $\Delta$ is
the site energy difference between sublattices, which also
corresponds to the band gap. From commutation relation
[$\Pi_{\pm},\Pi_{\mp}$]=$\mp2eB/\hbar$ where $\Pi_{\pm}=\Pi_{x}\pm
i\Pi_{y}$, we define ladder operators $b$ and $b^\dagger$ such
that: $\Pi_{+}=(\sqrt{2 eB/\hbar})b^{\dag}$ and $\Pi_{-}=(\sqrt{2
eB/\hbar})b$. The wave function is constructed from the linear
combination of eigenstates $| n\rangle$ of operator $b$. The
Hamiltonian (2) is diagonalized to yield the Landau levels, which
in conduction band is shown in Fig.1:
\begin{equation}
\varepsilon _n (\tau _z ) = \left\{ {\begin{array}{*{20}c}
   {\tau _z \frac{\Delta }{2}} \hfill & {(n = 0),} \hfill  \\
   { \pm \sqrt {\left( {\frac{\Delta }{2}} \right)^2  + \frac{{3a^2 t^2 eB}}{{2\hbar }}n} } \hfill & {(n = 1,2 \cdots ).} \hfill  \\
\end{array}} \right.
\end{equation}
where $\tau_z=\pm1$ labels the two valleys.

The magnetic susceptibility is defined as $ \chi \equiv  - \left(
{\frac{{\partial^2 F }}{{\partial B^2}}} \right)_\mu$, where the
thermodynamic potential can be calculated from the formula $F = -
\frac{1}{\beta}\frac{{eB}}{h}\sum\limits_{n,\tau_z} {\ln \left[ {1 +
e^{\beta (\mu - \varepsilon _n )} } \right]}$.  By using the
Euler-Maclaurin formula, we get the expand $F$ as a power series
with respect to $B$ and the analytical expression of susceptibility
is [10]
\begin{equation}
\chi(\mu)  = -\frac{{e^2 a^2 t^2 }}{{4\pi \hbar ^2 \Delta
}}\frac{{e^{\beta \mu } \left( {e^{\beta \Delta /2}  - e^{ - \beta
\Delta /2} } \right)}}{{1 + 2e^{\beta \mu } \textrm{Cosh}(\beta
\Delta /2) + e^{2\beta \mu } }}.
\end{equation}
Fig.2(a) shows the numerical result of the magnetic susceptibility
at $T=10$K. In the calculation we take $\Delta\sim0.28$ $eV$,
$t\sim2.82$ $eV$ and $a=2.46$\.{A}[5]. Compared with the zero gap
case, it is observed that the large diamagnetic dip is still visible
even when the gap opens, and it gets broadened in energy by the gap
width. Disorder effects smooth out this curve but will not change
the main features because susceptibility $\chi$ is a thermodynamic
property. We find that the integral of susceptibility over chemical
potential, i.e. $ \int {\chi (\mu )d\mu } $, is independent of both
gap size and temperature. At zero temperature, the susceptibility
becomes a square well shape, and vanishes when it is either
electron- or hole-doped.

The sudden jump of magnetic susceptibility at band edges signifies a
large paramagnetism from the carriers. Indeed, if we calculate the
magnetic susceptibility from Landau levels above the gap, then the
conduction band contribution to $\chi$ can be obtained as
\begin{equation}
\chi _C(\mu)  = \frac{{e^2 a^2 t^2 }}{{4\pi \hbar ^2 \Delta
}}\frac{1}{{1 + e^{ - \beta (\mu  - \Delta /2)} }}.
\end{equation}
From the plot in Fig.2(b), we can see that the large paramagnetic
response from the conduction electrons. What is the source of this
paramagnetism? Note that our model does not take into account of
spin degree of freedom, so it must come from the orbital motion of
electrons. In fact, previous studies have shown that there is an
intrinsic magnetic moment associated with the valley degree of
freedom[11]. At conduction band bottom ($\mu=\Delta/2$), the
moment equals the effective Bohr magneton
$\mu_B^*=\tau_z\frac{e\hbar}{2m_e^*}$ with the effective mass
$m_e^*=\frac{{2\hbar ^2 \Delta }}{{3a^2 t^2 }}$. If one models
this system by a 2D nonrelativistic electron gas with such a
pseudo-spin degree of freedom, one would obtain both Pauli
paramagnetic and Landau diamagnetic contributions recovering our
expression Eq(5) above.

Therefore, it is the valley magnetic moment that determines the
magnitude of the susceptibility signal. Using an experimentally
measured energy gap of 0.28 eV for graphene on SiC[5], the effective
mass is about 30 times smaller than the bare electron mass, hence
the magnetic moment is about thirty times larger than the free
electron spin magnetic moment.

\textit{Semiclassical Landau levels.} To further exemplify the
role of the valley magnetic moment, we now closely examine the
energy spectrum and analyze it via a semiclassical approach.  If
we place the exact Landau levels on the background of the original
conduction band dispersion curve, one discovers that Landau levels
from one valley start at the edge of the zero-field band (dashed
curves in Fig.1), while those from the other valley start from one
cyclotron energy above. The situation looks less odd if one shifts
the bands by the Zeeman energy $-\textbf{m}(\textbf{k})\cdot
\textbf{B}$, where $\textbf{m}(\textbf{k}) = \tau _z
\frac{{3e\Delta a^2 t^2 }}{{4\hbar \left( {\Delta ^2  + 3k^2 a^2
t^2 } \right)}}$ is the valley magnetic moment[12]. Relative to
the new bands (solid curves), the Landau levels of both valleys
now start at half of the cyclotron energy above the band edges,
which is the familiar result for 2D free electron gas.

Furthermore, one can obtain the Landau levels by semiclassically
quantizing the cyclotron orbits in these modified bands.  By taking
into account the Berry phase correction, Onsager's quantization
condition for the areas enclosed by the cyclotron orbits becomes
$\pi k^2 = \frac{{2\pi eB}}{\hbar }[n + \frac{1}{2} -
\frac{\Gamma(k) }{{2\pi }}]$[13]. In Ref.[12], the Berry curvature
of the conduction band at the two valleys has been calculated to be
$ \Omega(\textbf{k})  = \tau_z \frac{{3\Delta a^2 t^2 }}{{2(\Delta
^2 + 3k^2 a^2 t^2 )^{3/2} }}.$ The Berry phase $ \Gamma (k)$ can
then be obtained by integrating the Berry curvature over the area
enclosed by the orbit. The energies of the modified bands at these
quantized radius of the cyclotron orbits then yield the
semiclassical Landau levels.  Our result is shown in Fig.3, which
agrees very well with the exact Landau levels. However, we must
emphasize that for the low lying Landau levels close to the band
edge, the Berry phase effect is relatively less important than the
Zeeman shift associated with the valley magnetic moments.  The
situation reverses at high energies, where the magnetic moment
vanishes and the Berry phase approaches $\pi$.

\textit{Valley Polarization.} The valley magnetic moment allows a
direct coupling of the magnetic field with the valley degree of
freedom.  We expect that an equilibrium population difference
between the two valleys can be induced by applying a magnetic
field. Analogous to the spin polarization concept, we define
valley polarization to be $P_V=(n_{ + } - n_{ - } )/(n_{ + } + n_{
- } )$, where $n_{ \pm} $ is the densities of electrons in the
valley with index $\tau_z=\pm 1$. This can be calculated from the
relative number of occupied Landau levels in the two valleys,
resulting in the series of steps shown in Fig.4. Perfect valley
polarization occurs when there is one Landau level occupied in one
valley but none in the other(see Fig.1).

We have also calculated the valley polarization semiclassically, in
which the electron density from a given valley is obtained by
integrating the Berry-curvature modified density of states up to the
Fermi surface, i.e.   $\int^{\mu} {\frac{d \bf k}{{(2\pi )^2 }}}
\left( {1 + \frac{{e \bf B \cdot \bf \Omega  }}{\hbar }}
\right)$[14]. The result is plotted as the black curves in Fig.4,
and they smoothly go through the quantum steps. As expected, the
valley polarization increases with the magnetic field (Fig.4(a)),
and the induced polarization is largest at the band edge where the
valley magnetic moment is largest (Fig.4(b)).

\textit{Anomalous Hall effect. } It is now well established that in
the presence of an in-plane electric field, an electron will acquire
an anomalous velocity, proportional to the Berry curvature, in the
transverse direction [13]. This leads to an intrinsic contribution
to the Hall conductivity, $\sigma_{H}^{int}=2(e^2/\hbar) \int
\frac{d^2k}{(2\pi)^2}f(k)\Omega(k) $, where $f(k)$ is the
Fermi-Dirac distribution function, and the factor of 2 comes from
spin degeneracy. There is also a side-jump contribution [15]
proportional to the Berry curvature at the Fermi surface when
carriers scatter off impurities. Ignoring skew-scattering and other
effects due to inter-valley scattering, a valley-dependent Hall
conductivity at zero temperature is found in Ref.[12] as $ \sigma _H
(\tau _z ) = \tau _z \frac{{e^2 }}{h}\left[ {1 - \frac{\Delta
}{{2\mu }} - \frac{\Delta (4\mu ^2-\Delta^2)}{{8\mu ^3 }}} \right]
$. This result is independent of impurity density and strength. The
valley dependence in the Hall current will lead to an accumulation
of electrons on opposite sides of the sample with opposite valley
indices. Clearly, if there is a net valley polarization, a charge
Hall conductivity will appear upon the application of an electric
field, $ \sigma_{xy} = \frac{{e^2 }}{h}\frac{\Delta}{\bar \mu
^2}\left( \frac{3\Delta^2}{8\bar \mu^2}-1 \right)\delta \mu$. Here,
$ \bar \mu$ is the chemical potential in the absence of the magnetic
field, and $\delta \mu$ is the chemical potential difference between
the two valleys. We can express $\delta \mu$ in terms of the
field-induced valley polarization as $\delta \mu=\frac {\delta
\mu}{\delta n}(n_{+}+n_{-})\frac{\partial P_V}{\partial B}B=\frac
{2n}{D(\mu)}\frac{\partial P_V}{\partial B}B$, where $n$ is the
electron density and $D(\mu)$ is the density of states at the Fermi
level.

Assuming $\rho_{xx}\gg \rho_{xy}$, the Hall resistivity may be
expressed by $-\rho_{xy}=\rho_{xx}^2\sigma_{xy}=\gamma_{AH}B$,
where $\gamma_{AH}$ is the anomalous Hall coefficient[16] given by
\begin{equation} \gamma _{AH}  = \frac{1}{{\mu D(\mu )e}} \cdot
\frac{m_e}{{m_e^* }} \cdot \frac{\Delta }{{6\mu }} \cdot \left(
\frac{3\Delta^2}{8\mu^2}-1 \right)\cdot \left( {\frac{{e^2 }}{h}\rho
_{xx} } \right)^2,
\end{equation}
with $m_e$ being the bare electron mass. Figure 5 shows the
comparison of the anomalous Hall coefficient with the ordinary one
$\gamma _{OH}$ (=$-\frac{1}{ne}$) at zero temperature. It is shown
that with increase of chemical potential, $\gamma_{AH}$ changes from
positive to negative and has a curvature opposite to that of
$\gamma_{OH}$. Moreover, we find that $\gamma_{AH}$ is comparable
with $\gamma_{OH}$ for a typical value of $\rho_{xx}$=10
$k\Omega$[17] and becomes dominant for larger resistivity. It
indicates that Hall coefficient may be detected as a signal of Berry
curvature experimentally.

\textit{Valley-orbit coupling.} The traditional theories of
anomalous Hall effects are all based on two ingredients: spin
polarization and spin-orbit coupling [18].  In our system,
spin-orbit coupling is extremely weak [19], so we have ignored it
from the very beginning. Because of the magnetic moment associated
with the valleys, we are able to produce a population imbalance
between the two valleys, i.e., a valley polarization, by a
magnetic field.  Is there a valley-orbit coupling that underlies
the anomalous Hall effect discussed in this work?

From a phenomenological point of view, there is indeed a
valley-orbit coupling, because electrons in two valleys do have
opposite anomalous velocities under an electric field.  In fact,
the analogy with spin-orbit coupling can be made more precise if
we consider the effective Hamiltonian for the conduction band.
Ref.[13] has provided a recipe to construct the effective
Hamiltonian from three basic ingredients: the band energy, the
magnetic moment, and the Berry curvature.  The magnetic moment
gives rise to the Zeeman coupling between the magnetic field and
the valley index.  There is also a dipole-like term proportional
to the electric field $e\textbf{E} \cdot \textbf{R}$, where $\bf
R$ is the Berry connection potential defined by $ \nabla_k \times
\textbf{R} = \Omega $.  Using the expression of the Berry
curvature for the graphene system given by Eq.(2), we find that
$\textbf{R}=-\frac{\tau_z}{2k^2}\left( {1 - \frac{\Delta }{{\sqrt
{\Delta ^2 + 3k^2 a^2 t^2 } }}} \right)\textbf{k} \times
\bf{\hat{z}} $.  Therefore, the low-energy effective Hamiltonian
[20] can be written as
\begin{equation}
H_\text{eff}  = \varepsilon _0 (\textbf{k}) -\tau_z\mu^*_B
\textbf{B} \cdot \hat{\textbf{z}} - \frac{\tau_z
e\hbar^{2}}{4m_e^{*2}c^{*2}}  \hat{\textbf{z}} \cdot
(\textbf{k}\times \textbf{E}),
\end{equation}
where $c^*=\frac{\sqrt 3 at}{2\hbar}$ is the electron speed at
Dirac point of the gapless graphene. The third term is the
valley-orbit coupling, which has the similar form as the
spin-orbit coupling. The effective Hamiltonian resembles closely
the Pauli Hamiltonian for free electrons in the non-relativistic
limit.

In summary, we have studied the magnetic susceptibility and
anomalous Hall effect in epitaxial graphene. Our results show that a
large diamagnetism will result from the large valley magnetic moment
of graphene electrons. We also predict an anomalous Hall
coefficient, which can dominant that of the ordinary Hall effect,
due to a valley dependent Berry phase or valley-orbit coupling.

{\it Acknowledgments:} The authors acknowledge useful discussions
with D. Goldhaber-Gordon. This work is supported by NSF, DOE, the
Welch Foundation, and CNSF.

\newpage
\vskip 0.2cm \noindent {[1]K. S. Novoselov \textit{et al.}, Nature \textbf{438}, 197 (2005); Yuanbo Zhang \textit{et al.}, Nature \textbf{438}, 201 (2005)}\\
{[2]A. K. Geim and K. S. Novoselov, Nat. Mater. \textbf{6}, 183 (2007).}\\
{[3]Y. W. Son, M. L. Cohen and S. G. Louie, Phys. Rev. Lett. \textbf{97}, 216803 (2006).}\\
{[4]S. Y. Zhou \textit{et al.}, Nat. Mater. \textbf{6}, 770 (2007).}\\
{[5]A. Attausch and O. Pankratov, Phys. Rev. Lett. \textbf{99}, 076802 (2007); F. Varchon \textit{et al.}, Phys. Rev. Lett. \textbf{99}, 126805 (2007);J. Hass \textit{et al.}, Phys. Rev. Lett. \textbf{100}, 125504 (2008)}.\\
{[6]K. Novoselov, Nature Phys. \textbf{6}, 720 (2007).}\\
{[7]T. Ando, Physica E \textbf{40}, 213 (2007); J. W. McClure, Phys. Rev. \textbf{104}, 666 (1956).}\\
{[8]G. W. Semenoff, Phys. Rev. Lett. \textbf{53}, 2449 (1984).}\\
{[9]C. L. Kane and E. J. Mele, Phys. Rev. Lett. \textbf{95}, 226801 (2005).}\\
{[10]We also calculated the magnetic susceptibility from tight
binding model and find that the influence of valence band bottom on
susceptibility can be neglected. When the chemical potential is in
the gap, the susceptibility is calculated to be
$-4.935\times10^{-8}$ JT$^{-2}$m$^{-2}$, which is very close to our
analytical result $-5.084\times$10$^{-8}$ JT$^{-2}$m$^{-2}$ ($
\frac{{e^2 a^2 t^2 }}{{4\pi \hbar ^2\Delta
}}$). }\\
{[11]Wang Yao, Di Xiao, and Q. Niu, Phys. Rev.
B \textbf{77}, 235406 (2008).}\\
{[12]Di Xiao, Wang Yao, and Q. Niu, Phys. Rev.
Lett. \textbf{99}, 236809 (2007).}\\
{[13]M.-C. Chang and Q. Niu, Phys. Rev. B \textbf{53}, 7010 (1996).}\\
{[14]Di Xiao, Junren Shi and Q. Niu, Phys. Rev. Lett. \textbf{95},
137204 (2005).}\\
{[15]L Berger, Phys. Rev. B. \textbf{2}, 4559 (1970).}\\
{[16]D. Culcer, A. H. MacDonald, and Q. Niu, Phys. Rev. B \textbf{68}, 045327 (2003); J. Cumings \textit{et al.}, Phys. Rev. Lett. \textbf{96}, 196404 (2006).}\\
{[17]$\rho_{xx}\sim 4 k\Omega$ was observed in the gapless
graphene[1]. However, in the staggered-potential graphene, back
scattering process can not be prohibited completely, which leads
to the large $\rho_{xx}$. Moreover, the substrate plays an
important role on $\rho_{xx}$. For example, the resistivity of a
bilayer graphene sample was 10$^5$ $\Omega$ cm$^{-1}$ on a more
insulating 4H-SiC substrate compared with 0.2 $\Omega$ cm$^{-1}$
in 6H-SiC[5]. }\\
{[18]N. A. Sinitsyn, J. Phys:Condens. Matter\textbf{20}, 023201 (2008).}\\
{[19]H. Min \textit{et al.}, Phys. Rev. B \textbf{74}, 165310 (2006); Y. Yao \textit{et al.}, Phys. Rev. B \textbf{75}, 041401 (2007).}\\
{[20]M.-C. Chang and Q. Niu, J.Phys.:Condens. Matter \textbf{20}, 193202 (2008).}\\

\newpage
FIG.1: (Color online). The Landau levels from exact quantum
calculations for $\tau_z=+1$ (a) and $\tau_z=-1$ (b) valleys. For
comparison, the energy dispersions $\varepsilon_0(\textbf{k})$
(dashed line) and
$\varepsilon(\textbf{k})=\varepsilon_0(\textbf{k})-\textbf{m}(\textbf{k})\cdot
\textbf{B}$ (solid curves) are shown. Parameters are $\Delta$=
0.28 $eV$, $t$= 2.82 $eV$ and lattice constant $a$=2.46 \.{A}.

FIG.2: (Color online). (a) Magnetic susceptibility as a function
of chemical potential. (b) Magnetic susceptibility from conduction
electrons only. Here, $T=$ 10 $K$ and $\Delta$= 0.28 $eV$.

FIG.3: (Color online). The field dependence of semiclassical
Landau levels for $\tau_z=+1$ (red circle) and $\tau_z=-1$ (blue
triangle) valleys. For comparison, the exact quantum Landau levels
are also shown (solid line).

FIG.4: (Color online). (a)The variation of valley polarization
$P_V$ with the magnetic field $B$ ($\mu$=0.2 eV). (b)The variation
of valley polarization $P_V$ with the chemical potential $\mu$
(B=0.2 T). Both the quantum calculations (red) and semiclassical
results (black) are shown.

FIG.5: (Color online). The variation of anomalous Hall coefficient
$\gamma_{AH}$, ordinary Hall coefficient $\gamma_{OH}$ and the
total ($\gamma_{AH}$+$\gamma_{OH}$) with the chemical potential
$\mu$. Here, $\rho_{xx}=10$ $k\Omega$, $\Delta=0.28$ $eV$,
$t=2.82$ $eV$ and $a=2.46$\.{A}.

\newpage
\begin{figure}
\includegraphics[width=0.7\textwidth]{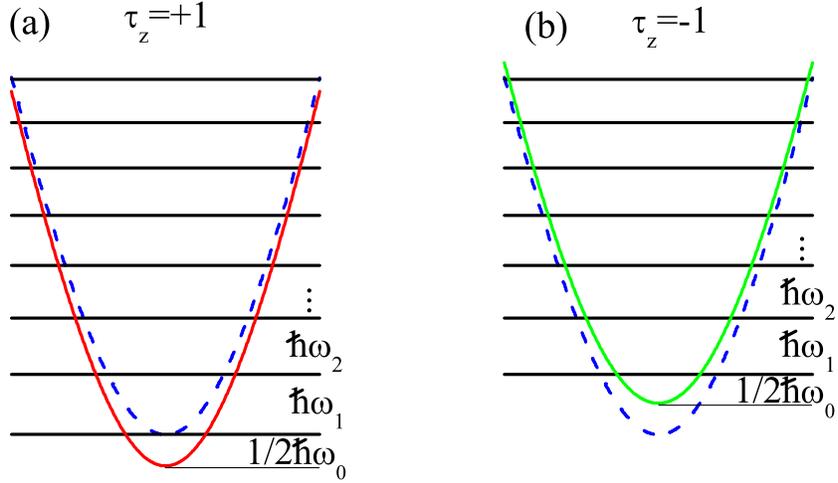}
\caption{\label{fig:epsart} (Color online).  The Landau levels
from exact quantum calculations for $\tau_z=+1$ (a) and
$\tau_z=-1$ (b) valleys. For comparison, the energy dispersions
$\varepsilon_0(\textbf{k})$ (dashed line) and
$\varepsilon(\textbf{k})=\varepsilon_0(\textbf{k})-\textbf{m}(\textbf{k})\cdot
\textbf{B}$ (solid curves) are shown. Parameters are $\Delta$=
0.28 $eV$, $t$= 2.82 $eV$ and lattice constant $a$=2.46 \.{A}. }
\end{figure}

\newpage
\begin{figure}
\includegraphics{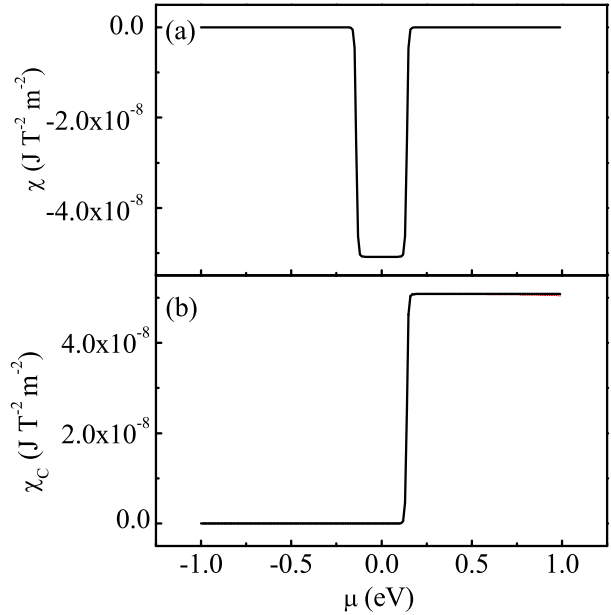}
\caption{\label{fig:epsart} (Color online).  (a) Magnetic
susceptibility as a function of chemical potential. (b) Magnetic
susceptibility from conduction electrons only. Here, $T=$ 10 $K$
and $\Delta$= 0.28 $eV$. }
\end{figure}

\newpage
\begin{figure}
\includegraphics[width=0.7\textwidth]{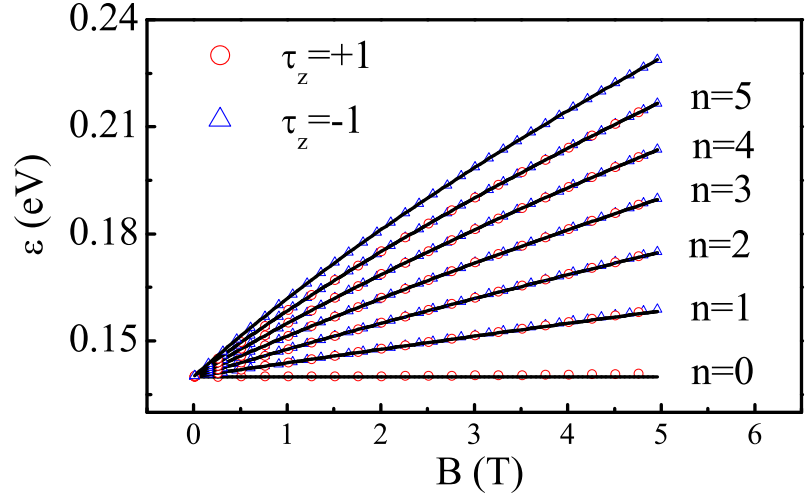}
\caption{\label{fig:epsart} (Color online). The field dependence
of semiclassical Landau levels for $\tau_z=+1$ (red circle) and
$\tau_z=-1$ (blue triangle) valleys. For comparison, the exact
quantum Landau levels are also shown (solid line). }
\end{figure}

\newpage
\begin{figure}
\includegraphics[width=0.9\textwidth]{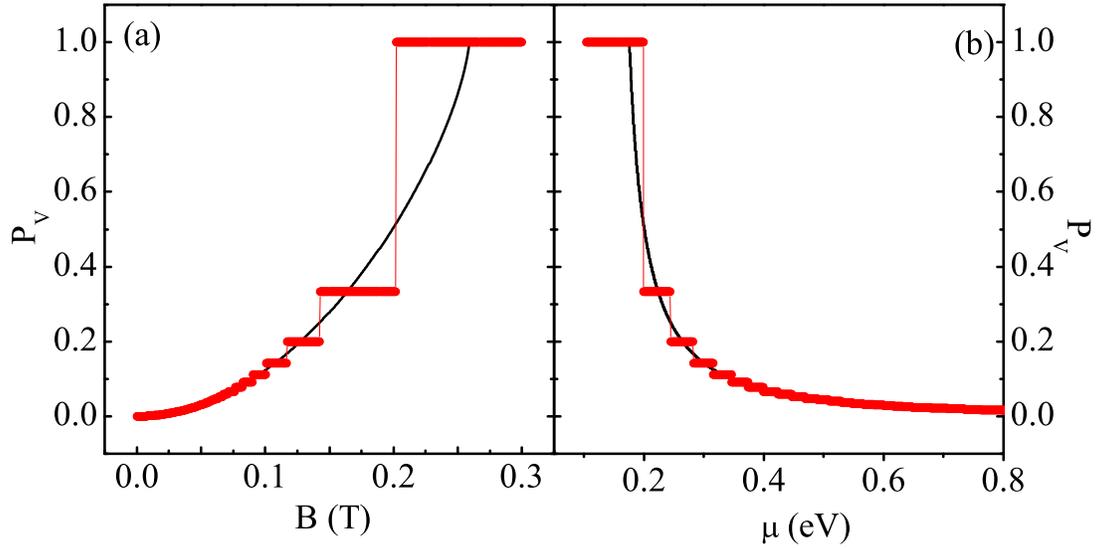}
\caption{\label{fig:epsart} (Color online). (a)The variation of
valley polarization $P_V$ with the magnetic field $B$ ($\mu$=0.2
eV). (b)The variation of valley polarization $P_V$ with the
chemical potential $\mu$ (B=0.2 T). Both the quantum calculations
(red) and semiclassical results (black) are shown. }
\end{figure}

\newpage
\begin{figure}
\includegraphics{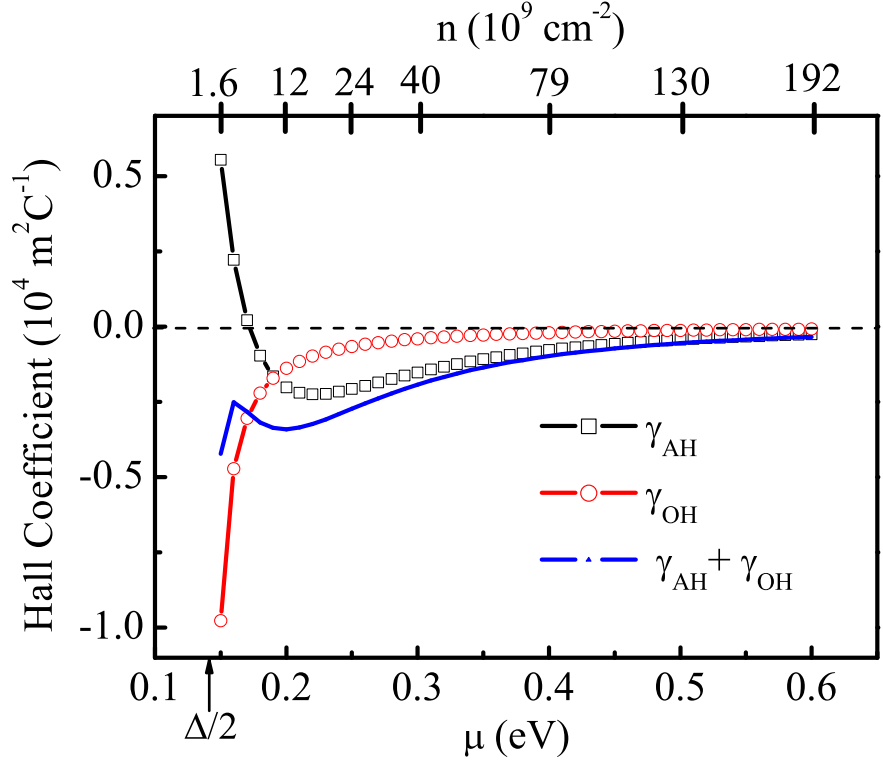}
\caption{\label{fig:epsart} (Color online). The variation of
anomalous Hall coefficient $\gamma_{AH}$, ordinary Hall
coefficient $\gamma_{OH}$ and the total
($\gamma_{AH}$+$\gamma_{OH}$) with the chemical potential $\mu$.
Here, $\rho_{xx}=10$ $k\Omega$, $\Delta=0.28$ $eV$, $t=2.82$ $eV$
and $a=2.46$\.{A}. }
\end{figure}
\end{document}